\documentclass[11pt]{article}
\usepackage{amsmath,amssymb,verbatim}
\usepackage[pdftex]{graphicx}
\def\abstract{\if@twocolumn
\section*{Abstract}
\else \normalsize
\begin{center}
{\bf Abstract\vspace{-.5em}\vspace{0pt}}
\end{center}
\quotation
\fi}
\def\endabstract{\if@twocolumn\else\endquotation\fi}
\newcommand{\bfw}{\mbox{\boldmath $w$}}

\newcommand{\bfg}{\mbox{\boldmath $g$}}

\newcommand{\bfv}{\mbox{\boldmath $v$}}
\newcommand{\bfV}{\mbox{\boldmath $V$}}

\newcommand{\fga}{\mbox{\boldmath $\alpha$}}
  \newcommand{\fgb}{\mbox{\boldmath $\beta$}}
  \newcommand{\fgd}{\mbox{\boldmath $\delta$}}
    \newcommand{\fgg}{\mbox{\boldmath $\gamma$}}

  \newcommand{\fgt}{\mbox{\boldmath $\tau$}}

\newcommand{\bq}{\begin{equation}}
\newcommand{\eq}{\end{equation}}
\newcommand{\bqq}{\begin{equation*}}
\newcommand{\eqq}{\end{equation*}}
\newcommand{\ben}{\begin{eqnarray}}
\newcommand{\een}{\end{eqnarray}}
\newcommand{\benn}{\begin{eqnarray*}}
\newcommand{\eenn}{\end{eqnarray*}}
\newcommand{\bi}{\begin{itemize}}
\newcommand{\ei}{\end{itemize}}

\newcommand{\bd}{\begin{description}}
\newcommand{\ed}{\end{description}}

\newcommand{\MCMC}{MCMC}
\newcommand{\MH}{MH}
\newcommand{\cdf}{cumulative distribution function}

\newcommand{\hatH }{{\widehat H}}
\newcommand{\hatPr }{{\widehat{\Pr}}}

\makeatletter
\@addtoreset{equation}{section}
\makeatother

\newtheorem{remark}{Remark}[section]

\newcommand{\myappendix}[1]{
    \renewcommand{\thesection}{{\bf Appendix}~{\bf \Alph{section}:}}
    \section{#1}
    \renewcommand{\thesection}{\Alph{section}}}

\date{\today}
\begin{document}

\title{\vspace{-1.0cm} Locally Adaptive Nonparametric Binary  Regression}
\author{$\hbox{SALLY A. WOOD, ROBERT KOHN and REMY COTTET}$\\
{\it University of New South Wales, Sydney, NSW, 2052, Australia}\\
 \and
$\hbox{WENXIN JIANG and MARTIN TANNER}$\\
{\it Northwestern University, Evanston, Illinois, 60208 U.S.A.}}
\maketitle

\begin{abstract} 
A nonparametric and locally adaptive Bayesian estimator is proposed for estimating a binary regression. Flexibility is obtained
by modeling the binary regression as a mixture of
probit regressions with the argument of each probit regression
having a thin plate spline prior with its own smoothing parameter
and with the mixture weights depending on the covariates. The estimator is 
compared to a single spline estimator and to a recently proposed locally 
adaptive estimator. The methodology is illustrated by applying it to both 
simulated and real examples. 
\end{abstract}

\section{Introduction} \label{s:introduction} 
Suppose we wish to model the spatial distribution of the habitat of the crested lark.  One way to do this is to model the probability of a crested lark sighting as a function of latitude ($lat$) and longitude ($lon$) as 
\begin{align} \label{e:intro prob} 
\Pr(\mbox{crested lark sighting}|lat,lon )& = H\left\{ g(lat,lon) \right\} , 
\end{align} 
where $H$ is a link function, such as a probit or logit, and $g$ is a function 
of latitude and longitude, which is either parametric or nonparametric. Our 
article presents a Bayesian method for estimating the probability in 
\eqref{e:intro prob} that does not assume a parametric form for $H$ and allows 
the probability to be locally adaptive with respect to the covariates, that 
is, to be smooth in one region of the covariate space and wiggly or even 
discontinuous in another. 

We model the binary regression as 
a mixture of probit binary regressions 
\begin{align} \label{e: mixt probit} 
\Pr(\text{crested lark sighting}|Lat,Lon ) & = \sum_{j=1}^r \pi_j(Lat,Lon) 
\Phi\left\{ g_j(Lat,Lon) \right\} \ , 
\end{align}
where $\Phi$ is the standard normal \cdf{} and the $g_j$ are truncated spline 
functions. We demonstrate that the resulting estimators are nonparametric and 
locally adaptive, but do not overfit. The reasons for this good performance 
are that mixing is done outside the probit \cdf{} rather than inside, the 
weights $\pi_j$ are allowed to vary with the covariates, and that the 
component functions $g_j$ can have different level of smoothness by having 
different smoothing parameters. This is especially important when modeling 
surfaces in two or more dimensions where a single smoothing parameter for a 
multidimensional surface will often be inadequate. The use of truncated spline 
bases allows models with several thousand observations and regression surfaces 
with a moderate number (at least 6 or 7) of covariates (provided the number of 
observations is adequate). Extending the methodology to higher dimensions 
problems is important because as the number of covariates increase so too does 
the need for local smoothing. This is because the smoothness of the function 
$H(x)$ is likely to be  different for different covariates.  
Our article  allows the number of components to 
vary from $r = 1, \dots, R$, with $R$ typically  3 or 4. We use a Bayesian approach and construct a Markov chain Monte Carlo sampling 
scheme (\MCMC) to estimate the model that uses the reversible jump method of 
Green~(1995) to move between model spaces having different numbers of components. 

Model~\eqref{e: mixt probit} is known as a Mixture-of-Experts (ME) model and 
was first introduced by Jacobs et al.~(1991) and Jordan and Jacobs~(1994), who 
used simple linear functions for the $g_j$ and estimated the model by the EM 
algorithm.  

There is an extensive literature on estimating binary regressions. 
McCullagh and Nelder~(1989) discuss parametric approaches. Nonparametric 
binary regression is discussed by Wang~(1994, 1997), 
Wahba, Wang, Gu, Klein and Klein~(1997) and Loader~(1999).  
Wood and Kohn~(1998) and  Holmes and Mallick~(2003) present Bayesian 
approaches to nonparametric binary regression. However, none of these
papers show that their estimates are locally adaptive. 
Kribovokova et al.~(2006) present an estimator of a binary regression that is 
based on quasi likelihood and show that it is locally adaptive.    

A number of locally adaptive estimators have recently been
proposed for Gaussian regression models. Most of these estimators
represent the unknown regression function as a linear combination
of basis functions. Frequentist approaches such as Friedman and
Silverman (1989), Friedman (1991) and Luo and Wahba (1997) sought
an optimal combination of basis functions using a greedy search
algorithm, whereas Bayesian approaches such as Smith and Kohn
(1996) and Denison, Mallick and Smith (1998) averaged over a large
number of combinations of subsets of the basis functions. 
\begin{comment}
Wood, Jiang and Tanner~(2002) obtained an adaptive estimator by mixing
over a combination of splines and used BIC to choose the number of 
components.  
\end{comment}

Wood, Jiang and Tanner~(2002) proposed a locally adaptive estimator for Gaussian regression by mixing over a combination of splines and used BIC to choose the number of components.  Our article builds on this Mixture of Experts approach by Wood,et al.~(2002). 

\begin{figure}[ht!]
\begin{center}
\includegraphics[scale=0.9]{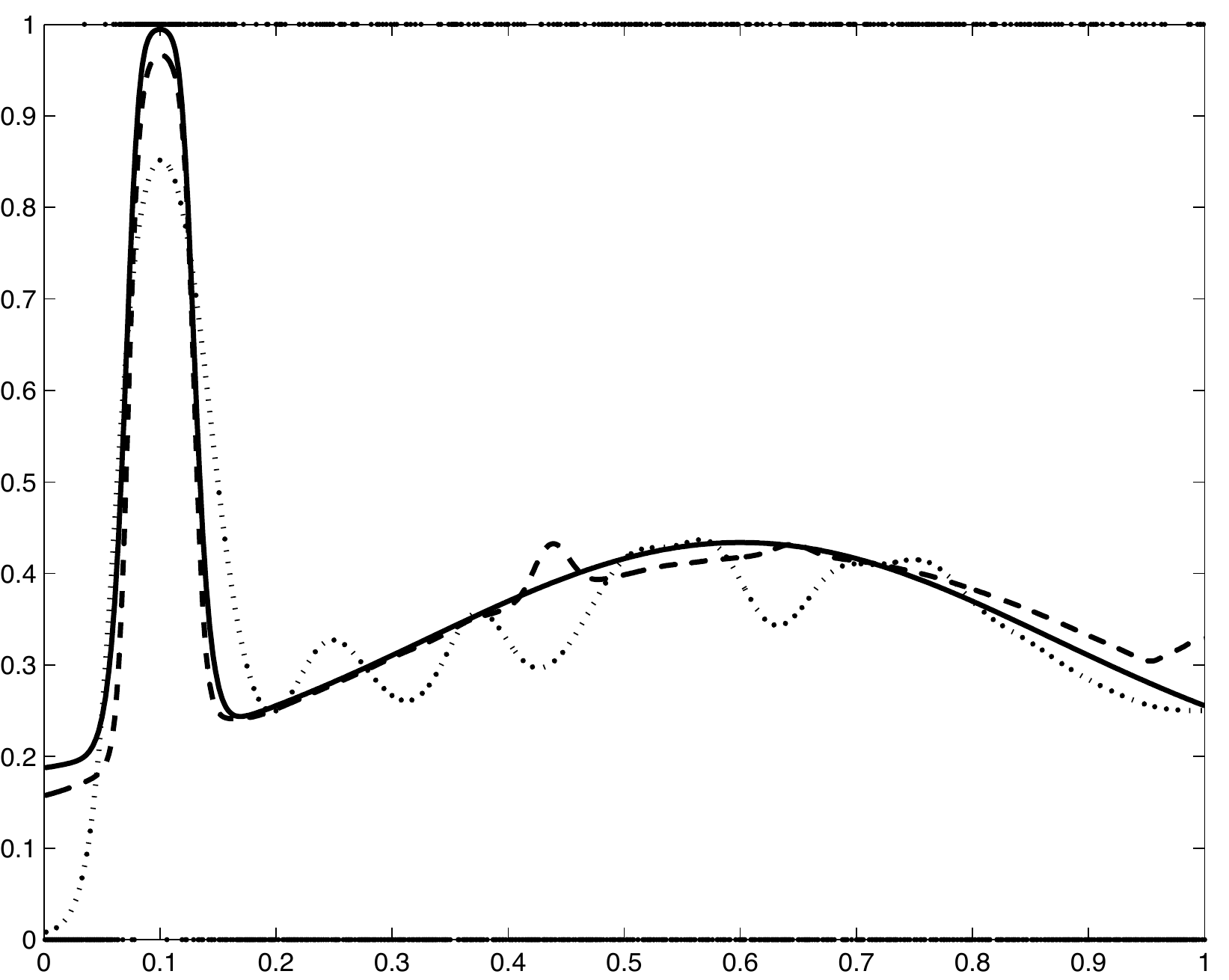} \caption{The
solid line gives the true function
$H(x)$ (Table \ref{table_1} function (b)). The
dotted line (...) is the estimate
$\hat{H}(x)$ for mixing on the inside while the dashed line (- - -)
 is the estimate $\hat{H}(x)$ for mixing on the outside. }
\label{fig_1.1}
\end{center}
\end{figure}

A direct way to obtain a locally adaptive estimator of binary 
probabilities is to adaptively estimate $g$ in (\ref{e:intro prob}) by using 
latent variables together with the basis selection methods in Friedman
(1991), Denison, Mallick and Smith (1998), and Smith and
Kohn~(1996) or with the mixture of splines method in Wood et
al.~(2002). However, we have found that for the mixture of
splines approach, adaptively estimating the regression function
$g$ by mixing on the inside of $\Phi$ results in poor estimates of
the probabilities due to overfitting. Figure~\ref{fig_1.1} 
gives an example of such overfitting. 
The figure  shows the true probability $H(x) =
Pr(y=1|x)$ and  the estimate $\widehat H(x) = \Phi (\widehat g(x)
) $ where $ g(x) =  \pi(x)  f_1 (x) + (1- \pi(x) )  f_2 (x) $,
which we call mixing on the inside.  The figure also shows the
estimate of $H(x)$ based on modeling $H(x)$ as $\pi(x) \Phi \{ f_1
(x) \} + (1-\pi(x) ) \Phi \{ f_2 (x) \}$, which we call mixing on
the outside. In this example, which is typical of all such
examples, it is clear that mixing on the inside does not perform
as well as mixing on the outside. The technical report by Wood, 
Kohn, Jiang and Tanner (2005) provides details
of why mixing on the inside tends to result in overfitting and
produces inferior estimates to mixing on the outside.

\section{Model and Prior Specification} \label{sect_2}
We present the model in this section. Appendix~A gives details of the sampling 
scheme used to estimate the model. 
Using the results in Tierney (1994) and Green (1995) we can show that the sampling scheme converges to the posterior distribution.

Let $w$ be a binary response variable taking the values 0 and 1. We model the 
binary regression of $w$ on $x$ by a mixture of finite but unknown number of 
probit regressions, as
\begin{align} \label{e: mixture} 
\Pr(w=1|x) & = \sum_{r=1}^R H_r(x) \Pr(r)\ , \text{   } H_r(x) = \sum_{j=1}^r 
\pi_{jr}(x) \Phi \{ g_{jr}(x) \} \\
\text{with} \quad \pi_{jr}(x) & = \exp(\delta_{jr}^\prime z)/ \sum_{k=1}^r \exp(\delta_{kr}^\prime z ) \ , j=1, \dots, r ,\nonumber 
\end{align} 
where $z = (1, x^\prime)^\prime $. 
We usually take the number of components $R$ as 3 or 4 with $\Pr(r) = 1/R$, for $r=1, \dots, R$. 
Without loss of generality we assume that the vector $\delta_{1r} = 0$ and let 
$\delta_r = (\delta_{2r}, \dots, \delta_{rr} )$ be a vector of unconstrained 
coefficients. 
We observe $w_1, \dots, w_n$ as well as the corresponding covariates $x_1, \dots, x_n$. 

To place a prior on $g_{jr}$ we write 
\begin{align} 
g_{jr} (x) & = \alpha_{jr}^\prime z + f_{jr}(x) 
\label{eqn_2.2.1}
\end{align}
where $\alpha_{jr}$ is a coefficient vector and $f_{jr}(x)$ is the nonlinear 
part of $g_{jr}$. For $j = 1, \dots, r$,  let  ${\bf{f}}_{jr}=\biggl ( f_{jr}(x_1), \dots, f_{jr}(x_n) \biggr ) ^\prime$. 
%, and let $x_{n+1}, \dots, x_{n}$ be additional points at which we may wish to estimate the binary regression. 
We write ${\bf{f}}_{jr}$ as a linear combination of basis functions as outlined below so that ${\bf{f}}_{jr} = X \beta_{jr}$, where the columns of the design matrix $X$ are partial thin plate spline basis functions and $\beta_{jr}$ is a vector of coefficients. 
%n $ m \times 1 $ vector of coefficients where $m$ is defined below. 
Appendix~B describes how we construct the design matrix $X$ to handle 
a large number of observations $n$ and a moderate number of covariates. 

The prior for $\alpha_{jr}$ for $j=1,\ldots,r$ is 
$N(0,c_{\alpha}I)$, for some large $c_{\alpha}$, and we
assume that the $g_{jr}$'s are independent {\em apriori}. Based on
emprical evidence we found that the regression function estimates
were insensitive to the choice of $c_{\alpha}$ over the range
$[10^2,10^{10}]$. The prior for $\beta_{jr} \sim N(0, \tau_{jr}I) , j=1, 
\dots, r$. We assume a uniform prior for $\tau_{1r} \sim U(0,c_{\tau})$, for
some large $c_{\tau}$ . To ensure identifiability we assume {\em
apriori} that $\tau_{jr} \sim U(0,\tau_{(j-1)r})$ for
$j=2,\ldots,r$, i.e., $\tau_{1r} <, \ldots, < \tau_{rr} <
c_{\tau}$.  The prior on the parameter vector $\delta_r $ for the mixing 
probabilities is $N(0, c_\delta I) )$, where $c_\delta = n$.

\section{Simulations}\label{sect_4} 
\subsection{Comparison with a single component estimator} \label{ss:single component}
The performance of the proposed method is studied
for four functions listed in table~\ref{table_1}, using a sample size of $n = 
1000$ for each function. 
The three univariate functions  are plotted 
in Figure~\ref{fig:univ_function_plots}. 
The bivariate function is plotted in Figure~\ref{fig_cylinder2exp}(a). 
The four functions 
are chosen in such a way that each requires a different type of smoothing. 
Function~(a) requires only one smoothing
parameter, function~(b) requires local smoothing, function~(c) is a 
discontinuous function, and function~(d) is a discontinuous bivariate function. 
\begin{table}[ht!]
\begin{center}
\begin{tabular}{|l|l|}
\hline 
Function Label & Function Formula \\ \hline
a (sin) & $\Pr(w=1|x)=\Phi \{2\sin (4\pi x)\}$ \\ 
b (peak)  & $\Pr(w=1|x)=\Phi \{\frac{5}{6}\exp {\left[ \frac{(x-0.1)^{2}}{0.18}\right]
}+\frac{1}{3}\exp {\left[ \frac{(x-0.6)^{2}}{0.004}\right] }-1\}$ \\ 
c (step)  & $\Pr(w=1|x)=\Phi \{-1.036+2.073I_{x}(0.25)-1.42712I_{x}(0.75)\}$ \\ 
& where $I_{x}(a)=1$ if $x>a$ \\ 
d (cylinder) & $\Pr(w=1|x)=D\{(x-0.5)^{2}+(y-0.5)^{2}-0.16^{2}\}$ \\ 
& where $D(x)=0.8$ if $x<0$ and $D(x)=0.2$ if $x>0$ \\ \hline
\end{tabular}
\end{center}
\caption{Regression functions used in simulations}
\label{table_1}
\end{table}
The estimates obtained using the proposed method are compared to
estimates obtained using a single component estimator as in Wood and
Kohn~(1998). We use this estimator for comparison because Wood and
Kohn~(1998) show that their single spline estimator outperforms
other available estimators, such as GRKPACK by Wang~(1997).
Fifty replications were generated for each regression function
 using a maximum $R = 3$ components. We use the average symmetric
Kullback-Leibler distance to measure performance.  This measure is
also used by Gu~(1992) and Wang~(1994), and is defined below. Let
\begin{eqnarray*}
I \{ H(x_i), \hatH(x_i) \} = {\hatPr} (w_i = 0 | x_i) \log
\left\{ \frac{{ \hatPr}  (w_i = 0 | x_i)} {Pr(w_i = 0 |
x_i)}\right\} + { \hatPr} (w_i = 1 | x_i) \log \left\{ \frac{
{ \hatPr}(w_i = 1 |x_i)} { \Pr(w_i = 1 | x_i)} \right\} \,.
\end{eqnarray*}
By Rao~(1973, pp.~58-59), $I \{ H(x_i), \hat H(x_i) \} \leq 0$ and
$I \{ H(x_i), {\hatH }(x_i) \} = 0$ if and only if $H(x_i) = {\hatH}
(x_i)$ at the design points. The average symmetric
Kullback-Leibler distance (ASKLD) between $H$ and $\hatH$ is
defined as
\begin{displaymath}
ASKLD( H, \hatH  ) = \frac{1}{n}\sum_{i=1}^n \left [I \{ H(x_i),
\hatH (x_i) \} + I\{\hatH (x_i) , H(x_i) \} \right ] \,.
\end{displaymath}
It follows from the properties of the Kullback-Leibler distance
that $ASKLD( H, \hatH ) \leq 0 $ and $ASKLD ( H, \hat {H} ) = 0
$ if and only if $\hatH  = H$. Thus, the closer $I ( H, \hatH
)$ is to zero the better.

\begin{figure}[htb!]
\begin{center}
\includegraphics[scale=0.6]{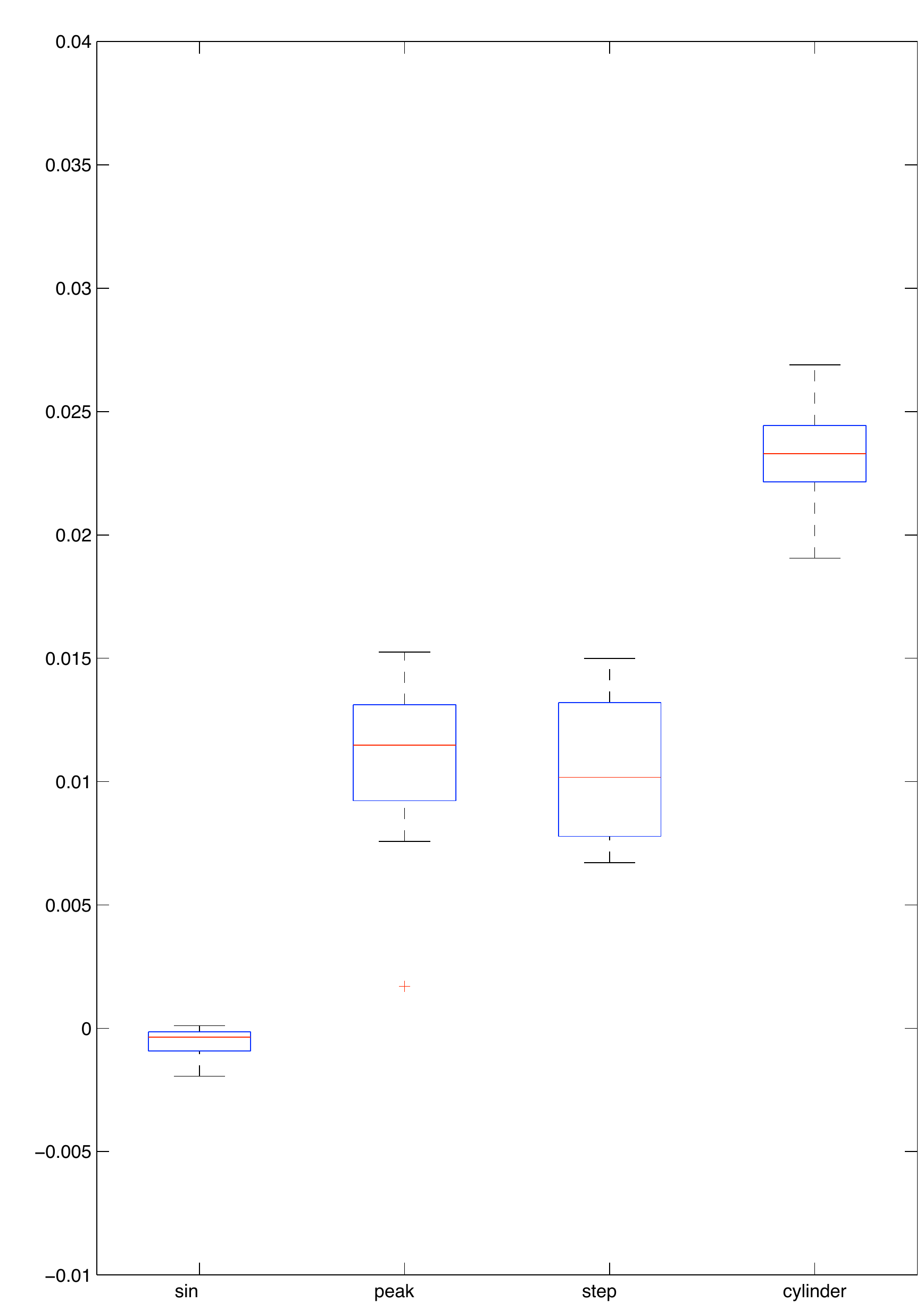}
\caption{Boxplots of the ASKLD between an estimate based on a mixture of
splines and an estimate based on a single spline for the functions~in table~
\ref{table_1}. Note that if the ASKLD$>0$ then the estimator
based on a mixture of splines is better than the estimator based on a single
spline.}
\label{fig1}
\end{center}
\end{figure}

\begin{figure}[ht!]
\begin{center}
\includegraphics[scale=0.85]{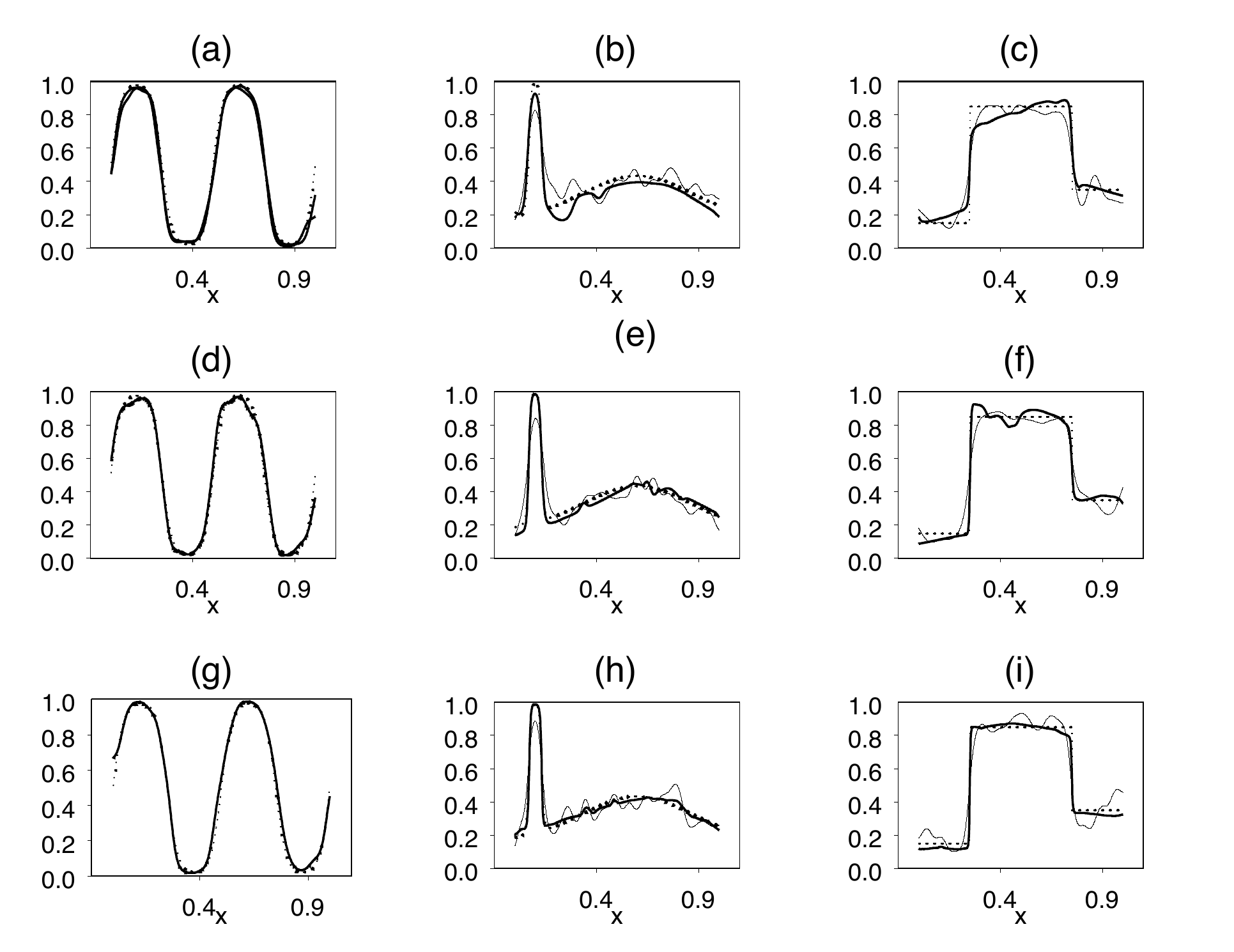} \caption{Panels~(a)--(c) plot
the estimates corresponding to the 10th worst percentile ASKLD for
functions~(a)~-~(c) in table~\ref{table_1}. Panels~(d)--(f) and
panels~(g)--(i) are similar plots corresponding to the 50th
percentile ASKLD and 10th best percentile ASKLD, respectively. In
all cases $n=1000$ and the true function $H(x)$ is given by the
dotted line, the estimate based on a mixture is given by the thick
solid line and the estimate based on a single spline is given by
the thin solid line.} 
\label{fig:univ_function_plots}
\end{center}
\end{figure}

\begin{figure}[bh]
\begin{center}
\includegraphics[scale=0.5]{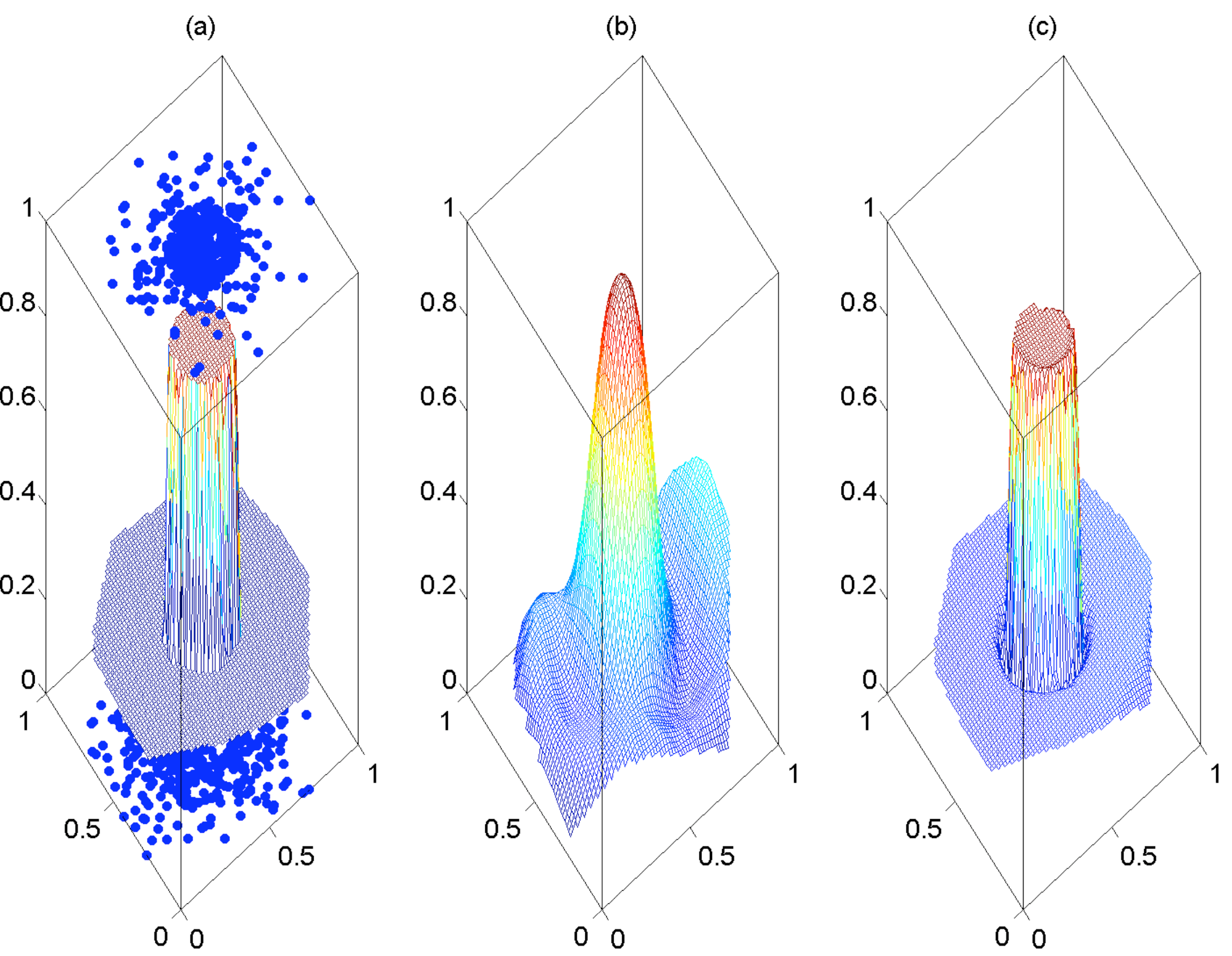}
\end{center}
\caption{Cylinder data,  panel (a) plots the true function and data, panel (b) plots the estimate for a single spline and panel (c) plots the estimate for a mixture of splines.}
\label{fig_cylinder2exp}
\end{figure}

For each replication, the ASKLD was calculated for both estimators,
where $\hatH(x_i)$ is the estimate obtained for the function
$Pr(w_i=1|x_i)$. Figure~\ref{fig1}  compares the
performance of the mixture estimator with the performance of the single 
component estimator using boxplots, 
with each boxplot representing the average 
difference in ASKLD for the fifty replications for that function.  
A positive difference means that the mixture model
performs better than the single spline estimator for that
replication, while a positive difference means the opposite.

\begin{comment}
\begin{figure}[tbh]
\begin{center}
\includegraphics[scale=0.6]{Boxplotsims1}
\caption{Boxplots of the ASKLD between an estimate based on a mixture of
splines and an estimate based on a single spline for the functions~in table~
\ref{table_1}. Note that if the ASKLD$>0$ then the estimator
based on a mixture of splines is better than the estimator based on a single
spline.}
\label{fig1}
\end{center}
\end{figure}
\end{comment}
Figure~\ref{fig1} shows that the mixture of splines estimator
performs significantly better in terms of the ASKLD than the
single spline estimator when the regression surface is
heterogenous, that is when it requires local smoothing.  
In particular, for the heterogeneous functions (b), (c) and (d) the mixture estimator outperformed the single spline.
 Furthermore, when the function is
homogenous, that is when it does not require local smoothing, 
the mixture estimator performs almost as well as a single
spline estimator because when only one spline is needed, for
example for function~(a), the posterior probability of a single
spline is high.

Table~\ref{table_2} gives the average (over the 50 replications) 
 posterior probability of the
number of splines needed for mixing for each of the four
regression functions. The table shows that for the homogenous function (a) 
the average posterior probability that only one spline is need is
0.78. In general we have found that for homogenous functions the posterior 
probability of a single component is high. Conversely, we have found that 
for heterogenous functions the posterior probability of requiring more than a 
single component is high. Thus, for function~(b), the average posterior 
probability is 0.76 that two splines are needed and 0.24 that three spline are 
needed. For the piecewise constant functions~(c)~and~(d), the average
posterior probabilities that two splines are needed are 0.60 and 0.84 
respectively, whereas the average posterior probabilities that three splines are need is 0.37 and 0.08 respectively. 

\begin{table}[ht!]
\begin{center}
\begin{tabular}{|l|r|r|r|}
\hline 
Function Label & \multicolumn{3}{c|}{$\Pr(r|\bfw)$} \\
\hline
& $1$ & $2$ & $3$ \\ \hline
a (sin) & $0.78$ & $0.21$ & $0.01$ \\ 
b (peak) & $0.00$ & $0.76$ & $0.24$  \\ 
c (step) & $0.03$ & $0.60$ & $0.37$\\
d (cylinder) & $0.08$ & $0.84$ & $0.08$
 \\ \hline
\end{tabular}
\end{center}
\caption{Average posterior probability of number of splines needed}
\label{table_2}
\end{table}

To see how the difference in ASKLD translates into differences in
the regression function estimates, the estimates  for functions~(a)-(c)  corresponding to
the 10th percentile, 50th percentile and 90th percentile ASKLD are plotted in Figure~\ref{fig:univ_function_plots}.  Figure~\ref{fig:univ_function_plots} shows that an estimator based on a
mixture of splines can accurately estimate both homogeneous and
heterogeneous functions. When the function is homogeneous, e.g.
function (a), the mixture estimates are visually indistinguishable
from the single spline estimates. However, when the function is
heterogeneous, the single spline estimates are much worse than the
mixture estimates. For example, even the $90^{th}$ percentile of
the single spline estimate of function (b) fails to capture the
peak on the left and overfits on the right, whereas the mixture
estimate does well throughout the whole range of the function. 
Figure \ref{fig_cylinder2exp} plots the fit for the cylinder data for a single spline and 
 a mixture of splines. The improvement is marked,
with the single spline being unable to capture the sudden change of
curvature. 
\begin{comment}
\begin{figure}[hb]
\begin{center}
\includegraphics[scale=0.85]{GS10} \caption{Panels~(a)--(c) plot
the estimates corresponding to the 10th worst percentile ASKLD for
functions~(a)~-~(c) in table~\ref{table_1}. Panels~(d)--(f) and
panels~(g)--(i) are similar plots corresponding to the 50th
percentile ASKLD and 10th best percentile ASKLD, respectively. In
all cases $n=1000$ and the true function $H(x)$ is given by the
dotted line, the estimate based on a mixture is given by the thick
solid line and the estimate based on a single spline is given by
the thin solid line.} \label{fig4}
\end{center}
\end{figure}

\begin{figure}[bh]
\begin{center}
\includegraphics[scale=0.5]{TrueAndFitRJ}
\end{center}
\caption{Cylinder data,  panel (a) plots the true function and data, panel (b) plots the estimate for a single spline and panel (c) plots the estimate for a mixture of splines.}
\label{fig_cylinder2exp}
\end{figure}
\end{comment}

\subsection{Comparison to a locally adaptive estimator} \label{ss:adaptfit}
Kribovokova et al.~(2006) propose a general approach for locally adaptive 
smoothing in regression models with the regression function modeled as a 
penalized spline that has a smoothly varying smoothing parameter function 
which is also modeled as a penalized spline. 
In particular, their approach handles local smoothing of 
binary data. The authors have implemented their approach in a package called 
Adaptfit which is written in R and is available at 
http://cran.r-project.org/src/contrib/Descriptions/AdaptFit.html. 
This section compares our approach to that of Kribovokova et al.~(2006) as 
implemented in Adaptfit in terms average squared error, coverage probabilities 
and running time for the four functions (a)--(d). The results reported 
below are each based on 50 replications unless stated otherwise. 

Let $\text{ASE}_{ME}$ be the average squared error over the abcissae of the 
data of the Bayesian ME estimator and let $\text{ASE}_{AF}$ be the 
corresponding averaged squared error of the Adaptfit estimator. We define the 
percentage change in going from the ME estimator to the adaptfit estimator as 
\begin{align*}
\% \Delta  \text{ASE} & = 
\frac{ ASE_{AF} - ASE_{ME}}{ASE_{ME}} \times 100 \ . 
\end{align*}
Table~\ref{t:avse} reports the 25th, 50th, 75th percentiles and the mean 
of $\% \Delta  \text{ASE}$ and shows that the ME estimator and the 
Adaptfit estimator perform similarly for functions (a) and (b), but that the 
ME of experts estimator outperforms the Adaptfit estimator 
for functions (c) and (d). 

\begin{figure}[ht!]
\centering
\includegraphics[angle=0, width=1.0\textwidth]{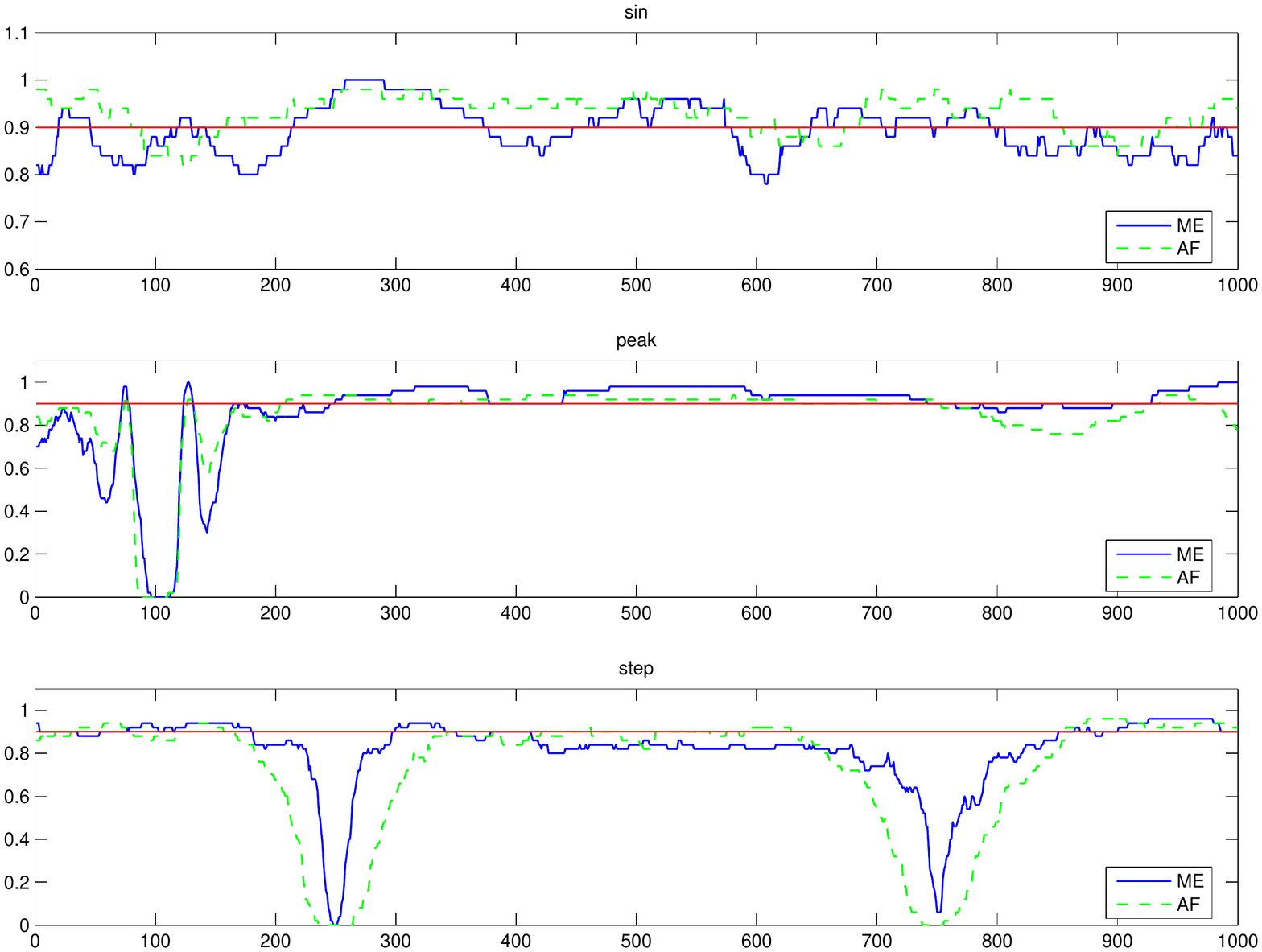}
\caption{Plots of the pointwise empirical coverage probabilities for the mixture of experts (ME) and adaptive fit (AF) estimators when the nominal coverage probability is 0.9. The plots are for the functions (a)--(c).}
\label{fig:ecp}
\end{figure}

\begin{figure}[ht!]
\centering
\includegraphics[angle=0, width=1.0\textwidth]{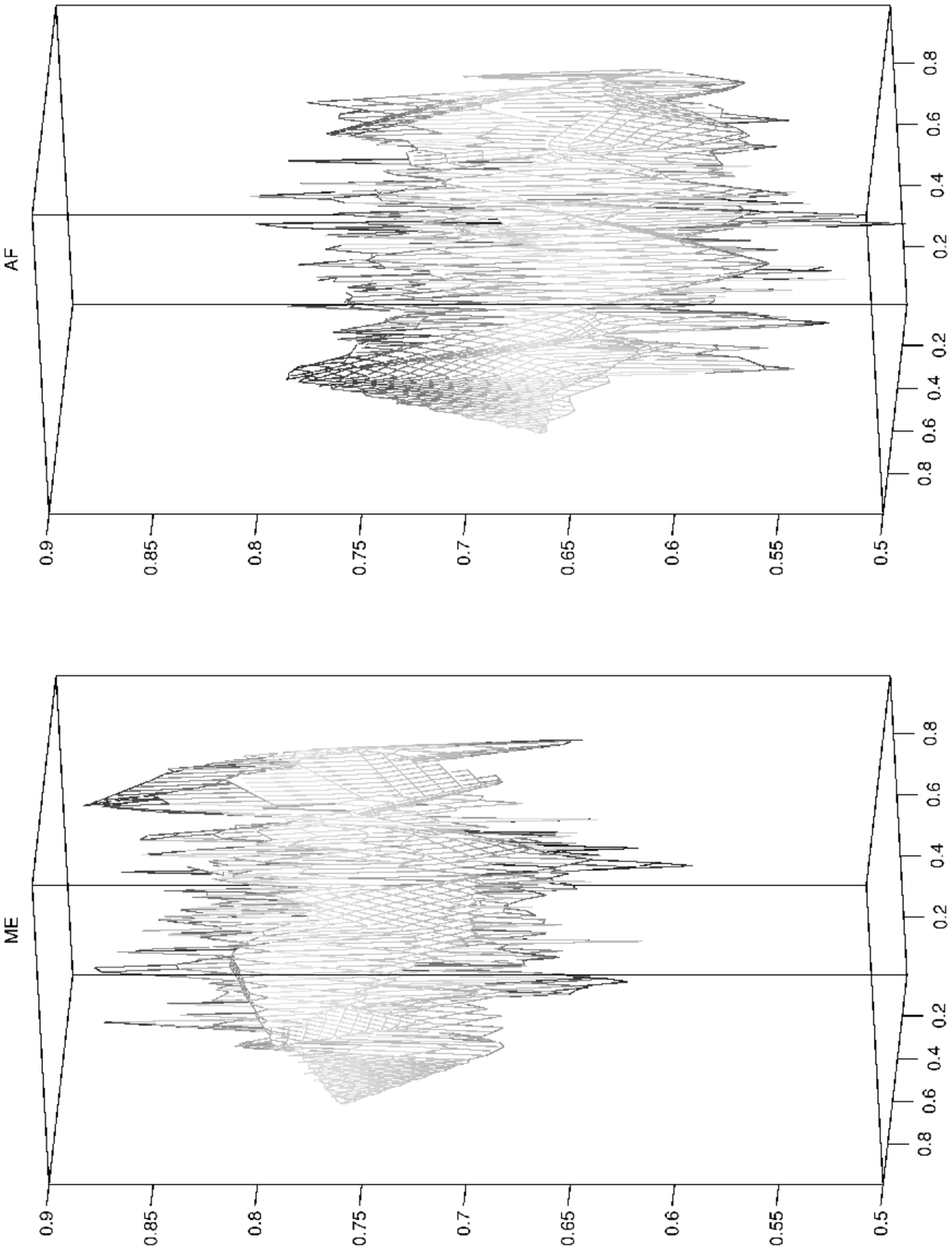}
\caption{Plots of the pointwise empirical coverage probabilities for the mixture of experts (ME) and adaptive fit (AF) estimators when the nominal coverage probability is 0.9. The plots are for the function (d).}
\label{fig:ecp_d}
\end{figure}

Let the empirical coverage probability 
$\text{ECP}_{ME}(x)$ be the proportion (out of 50) 
of 90\% pointwise confidence 
intervals that contain the true probability at the abcissa $x$ for the ME 
estimator. Let $\text{ECP}_{AF}(x)$ be defined similarly for the Adaptfit 
estimator. Figure~\ref{fig:ecp} plots $\text{ECP}_{ME}(x_i)$ and 
$\text{ECP}_{AF}(x_i)$ at the  abcissae $x_i$ of the data for the functions 
(a)--(c). Figure~\ref{fig:ecp_d} is a similar plot for the cylinder function (d). Let 
\begin{align*}
\% \Delta \text{AECP}_{ME} & = 100 \times \biggl ( n^{-1} \sum_{i=1}^n 
\text{ECP}_{ME}(x_i)-0.9 \biggr ) / 0.9 
\end{align*}
be the percentage deviation from 0.9 of the average of the 
$\text{ECP}_{ME}(x_i)$ over 
all the  $x_i$ abcissae of the data. 
Let $\% \Delta \text{AECP}_{AF}$ be defined similarly for 
the Adaptfit estimator. Table~\ref{t:avse} reports both $\% \Delta \text{AECP}_{ME}$ and $\% \Delta \text{AECP}_{AF}$. 
Figure~\ref{fig:ecp} and table~\ref{t:avse} suggest that the 
empirical coverage probabilities of the ME estimator and Adaptfit are similar 
for functions (a) and (b), while the ME estimator has superior empirical 
coverage probabilities for functions (c) and (d). 

\begin{table}[ht!] 
\begin{center} 
\begin{tabular}{|l|cccc|cc|} \hline 
Function & \multicolumn{4}{|c}{$\% \Delta  \text{ASE}$} & 
\multicolumn{2}{c|}{$\% \Delta \text{AECP}$}  \\
\hline
 & 25 & 50 & 75 & mean & ME &AF   \\ \hline 
(a) Sin & $-11.32$ & $7.86$ & $29.64$ & $11.18$ & $-0.74$ & $3.10$  \\
(b) Peaks & $-8.03$ & $-0.703$ & $15.07$ & $7.72$& $-3.34 $ & $-5.89 $ \\
(c) Step & $52.23$ & $93.15$ & $140.23$ & $101.99$& $-9.07 $& $-18.12$   \\
(d) Cylinder & $86.28 $ & $130.76$ & $162.11$ & $126.83$ & $-16.80 $ & $-26.62$ 
\\ \hline 
\end{tabular} 
\end{center}
\caption{Comparison of Bayesian ME and adaptfit. The first four  columns give 
the 25th, 50th , 75th percentiles and the mean 
of the percentage difference between the 
averaged mean squared error of adaptfit and the ME estimators. The next two 
columns gives the percentage coverage errors of the 90\% confidence intervals 
for ME and Adaptfit.}
\label{t:avse}
\end{table}

We now discuss some computational issues that determine  the  
performance of Adaptfit in the binary regression case. 
Adaptfit uses two sets of knots. The first set 
of knots is for the penalized spline basis for the regression function and 
the second set of knots is for the penalized spline basis for the smoothing 
parameters. See Kribovokova et al.~(2006) for details. 
Let $K_b$ be the number of knots chosen for the first penalized 
spline and $K_c$ the number of knots chosen for the second penalized 
spline. We found in the binary case that 
if a function requires a locally adaptive estimator then to get a satisfactory 
fit it may be necessary to take $K_b$ quite large. 
For example, for the peak 
function (b) and the cylinder function (d) 
it seems necessary to take $K_b=120$ to 150. 
The results for functions (a), (c) 
and (d) in table~\ref{t:avse} were obtained using using $K_b = 150$ and $K_c=20$. The results for  function~(b) (peak) were supplied to us by 
Dr Tatyana Krivobokova who used an 
Splus implementation of Adaptfit because we were unsuccessful with the R 
implementation . The results reported 
for Adaptfit for function (b) are for 47 replicates with the other three 
replicates being unsatisfactory. Adaptfit also has a 
default choice of $K_b$ and $K_c$ and we checked that the above settings for 
functions (a), (c) and (d) gave 
as good or better performance as the defaults. The times given below are 
averages over several replications and were obtained on a 2.8GHz PC running 
Matlab 7. For function (a), 
Adaptfit took 25 seconds for the default number of knots and 320 seconds for 
$(K_b,K_c)=(150,20)$. For function (c), Adaptfit took 36 seconds for 
the default number of knots and 325 seconds for $(K_b,K_c)=(150,20)$. For 
function (d), Adaptfit did not give satisfactory results for the 
default number of knots and took 316 seconds for $(K_b,K_c)=(150,20)$. The ME 
estimator takes about 90 seconds per 2000 iterations for each of the examples 
in this section. 
For each of the examples reported in this section and the next section
we ran the ME estimator for 10000 iterations in total 
(5000 warmup and 5000 sampling), that is, the time taken for each of the 
examples in this section is about 450 seconds. However, we have 
found in extensive testing that it is sufficient to use 4000 to 6000 
iterations in all of these examples, that is about 180 to 
270 seconds in total. All the times reported 
were recorded on a 2.8GHz PC running Matlab 7. Thus Adaptfit can be 
very fast when the default number of knots is used, but for a binary regression 
it is difficult to determine apriori for any data set whether the default 
number of knots is adequate and in our opinion it is safer to take the more  
conservative approach by setting $K_b$ to 120 or 150. In that case the times 
required by Adaptfit and the ME estimators are not that different, while the 
ME estimator in the binary case appears computationally more robust. 

\begin{comment}
\begin{figure}[ht!]
\centering
\includegraphics[angle=0, width=1.0\textwidth]{CoverageMEuniv.pdf}
\caption{Plots of the pointwise empirical coverage probabilities for the mixture of experts (ME) and adaptive fit (AF) estimators when the nominal coverage probability is 0.9. The plots are for the functions (a)--(c).}
\label{fig:ecp}
\end{figure}

\begin{figure}[ht!]
\centering
\includegraphics[angle=0, width=1.0\textwidth]{CoverageMEmv.pdf}
\caption{Plots of the pointwise empirical coverage probabilities for the mixture of experts (ME) and adaptive fit (AF) estimators when the nominal coverage probability is 0.9. The plots are for the function (d).}
\label{fig:ecp_d}
\end{figure}
\end{comment}

\section{Real Examples} \label{s:examples}
\subsection{Probability of crested lark sighting in Portugal}\label{ss:lark} 

This section demonstrates how the proposed method can be used to model spatial data by modelling  the probability of a crested lark sighting at various locations in Portugual.  The data were obtained from Wood (2006).   Each observation refers to one tetrad (2km by 2km square) and contains a variable indicating whether the crested lark was sighted in the tetrad  or not together with the location of the tetrad.  The location of a tetrad is identified by kilometers east and north of an origin. Portugal can be divided into 25100 tetrads for which there were observations on 6457 tetrads.  This dataset was analysed by Wood  (2006) who aggregated the data into 10km by 10km squares and fitted a binomial generalized additive model (GAM) using thin plate regression splines to the aggregated response. In their example the degree of smoothness was estimated using the un-biased risk estimator (UBRE), which was then scaled-up by a factor of 1.4 to avoid overfitting. The factor by which the smoothing parameter is rescaled is chosen subjectively and affects the estimated probabilities substantially.  In contrast, our method allows for the degree of smoothness to vary across the covariate space and the estimated probabilities are therefore spatially adaptive . 
 
  Our model for the probability of a crested lark sighting is given by (\ref{e: mixture}) and (\ref{eqn_2.2.1}) with $x=(east,north)$, where $east$ and $north$ mean kilometers east and north of an origin. The dependent variable is 1 if a crested lark is sighted in a tetrad and 0 if it is not. We assume a maximum of four mixture components. 
  
\begin{figure}[bh]
\begin{center}
\includegraphics[scale=1]{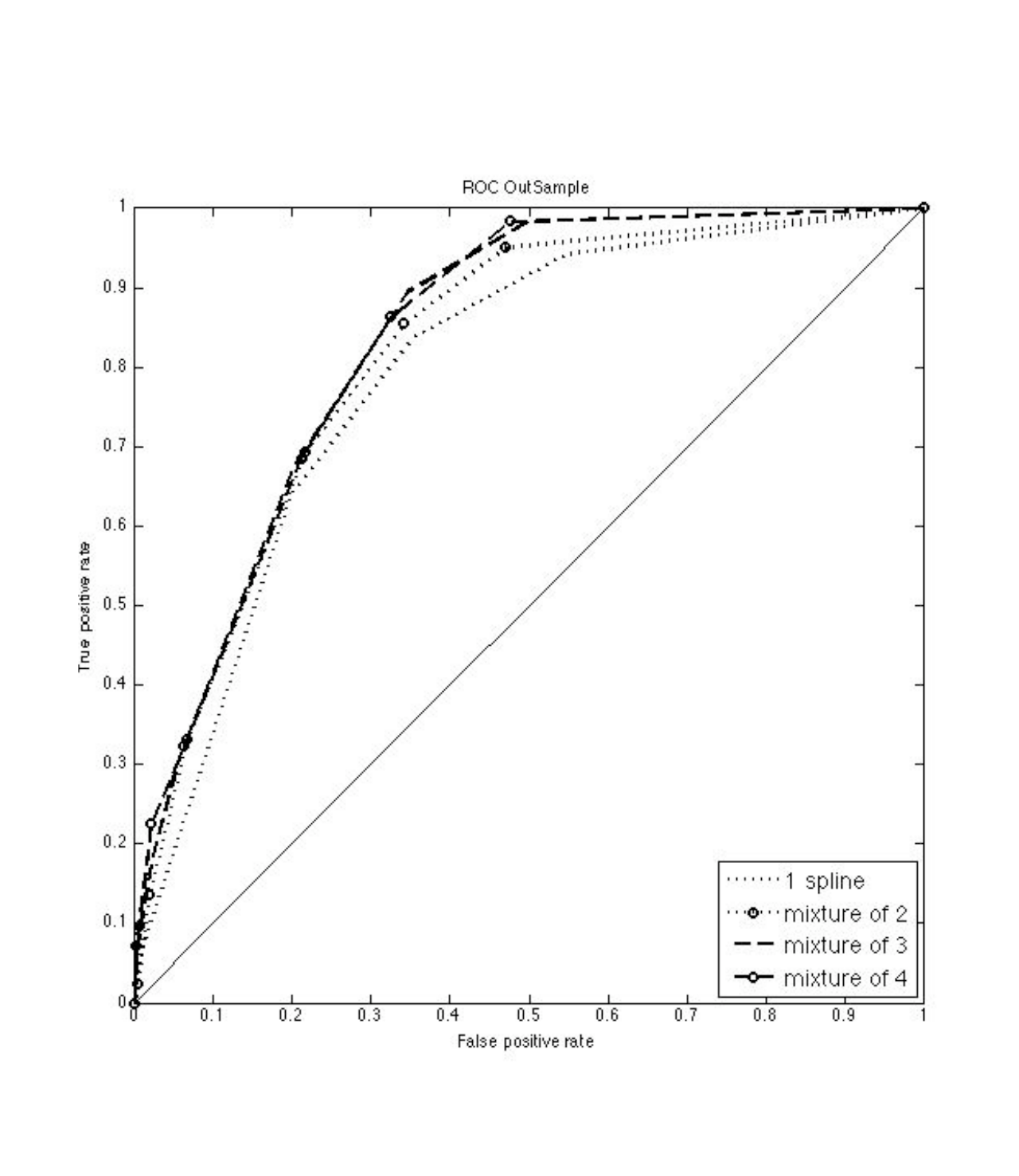}
\end{center}
\caption{ROC for out of sample data (data) for different number of mixture components.}
\label{fig_roc}
\end{figure}

We considered the choice of the number of mixture components in several ways. 
First, the posterior probabilities for 1 to 4 components are 0.0, 0.94, 0.04 and 0.01, suggesting a two component mixture. We also looked at the empirical 
receiver operating curves (ROC) for models with 1 to 4 components. 
See Fan,  Upadhye and Worster (2006) for a description of ROC curves. In our 
case a ROC curve shows the trade off between classifying a tetrad as 
containing the crested lark when the 
tetrad does contain the crested lark versus classifying a tetrad as containing 
the crested lark when it does not.   The larger the area under the ROC curve 
the more effective is the model at classifying. Out of 6457 observations we 
randomly selected 5817 for model fitting and set aside 640 for model testing.  
We estimated the probability that a tetrad contains the crested lark using 
mixture models containing one to four components. For each mixture model we 
then used the estimated probabilities to classify the 640 observations set 
aside for model testing.  Figure~\ref{fig_roc} plots the empirical ROC curves 
based on these 640 observations and shows the improvement in classification 
that is achieved by using a mixture of two splines over a single spline. 
Figure~\ref{fig_roc} also shows that there is not much improvement in 
classification in moving from a mixture of two to a mixture of three or four 
and hence supports the choice of a mixture of two splines as the preferred 
model. This finding is consistent with the posterior probabilities of the 
number of components. 
 
\begin{figure}[bh]
\begin{center}
\includegraphics[scale=.5]{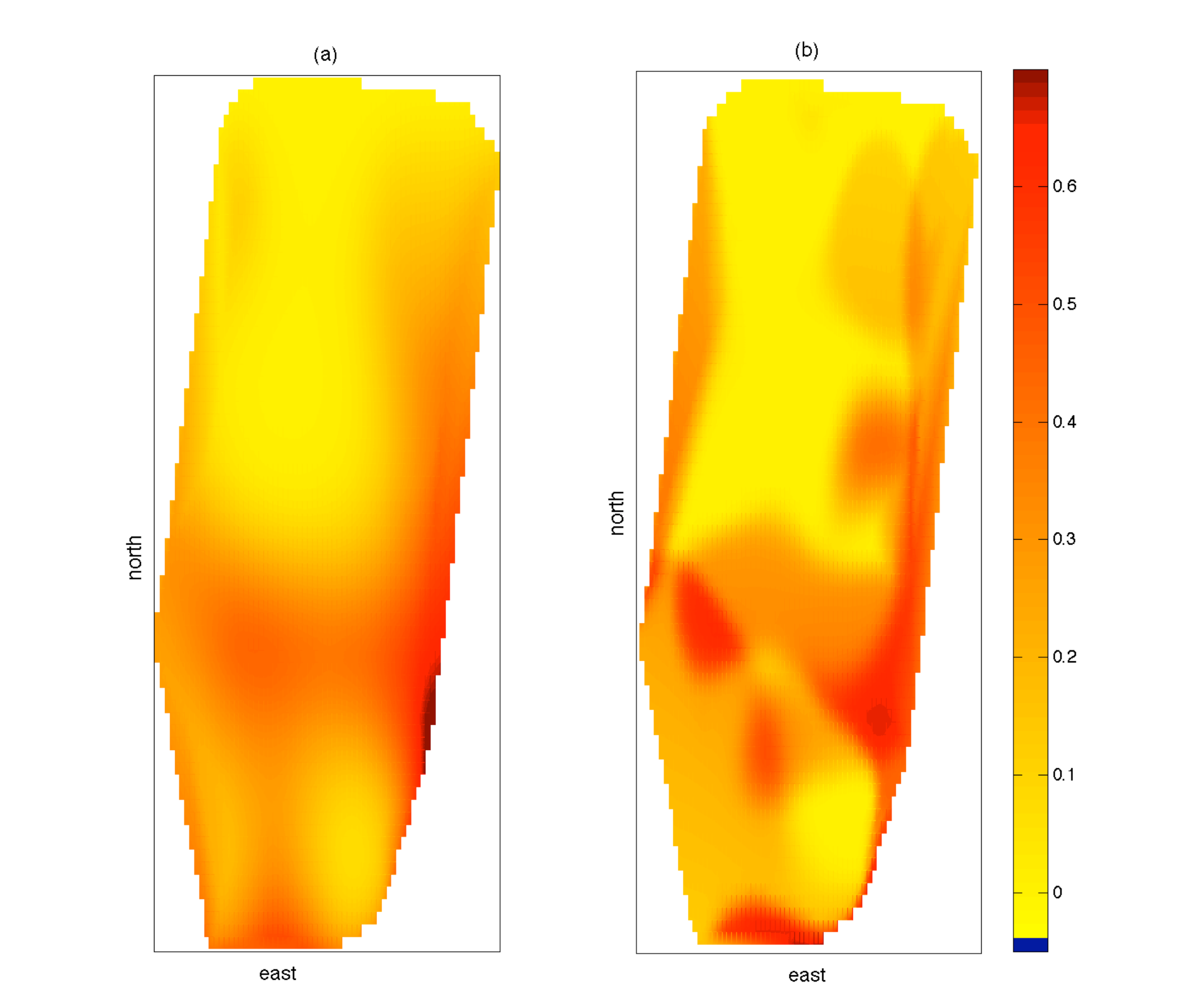}
\end{center}
\caption{Contour plot for $\Pr (\mbox{crested lark sighting}=1|east, north)$ for a single spline estimator, panel (a) and a mixture of splines estimator panel (b).}
\label{fig_birds}
\end{figure}

\begin{figure}[htbp]
\begin{center}
\includegraphics[scale=.8]{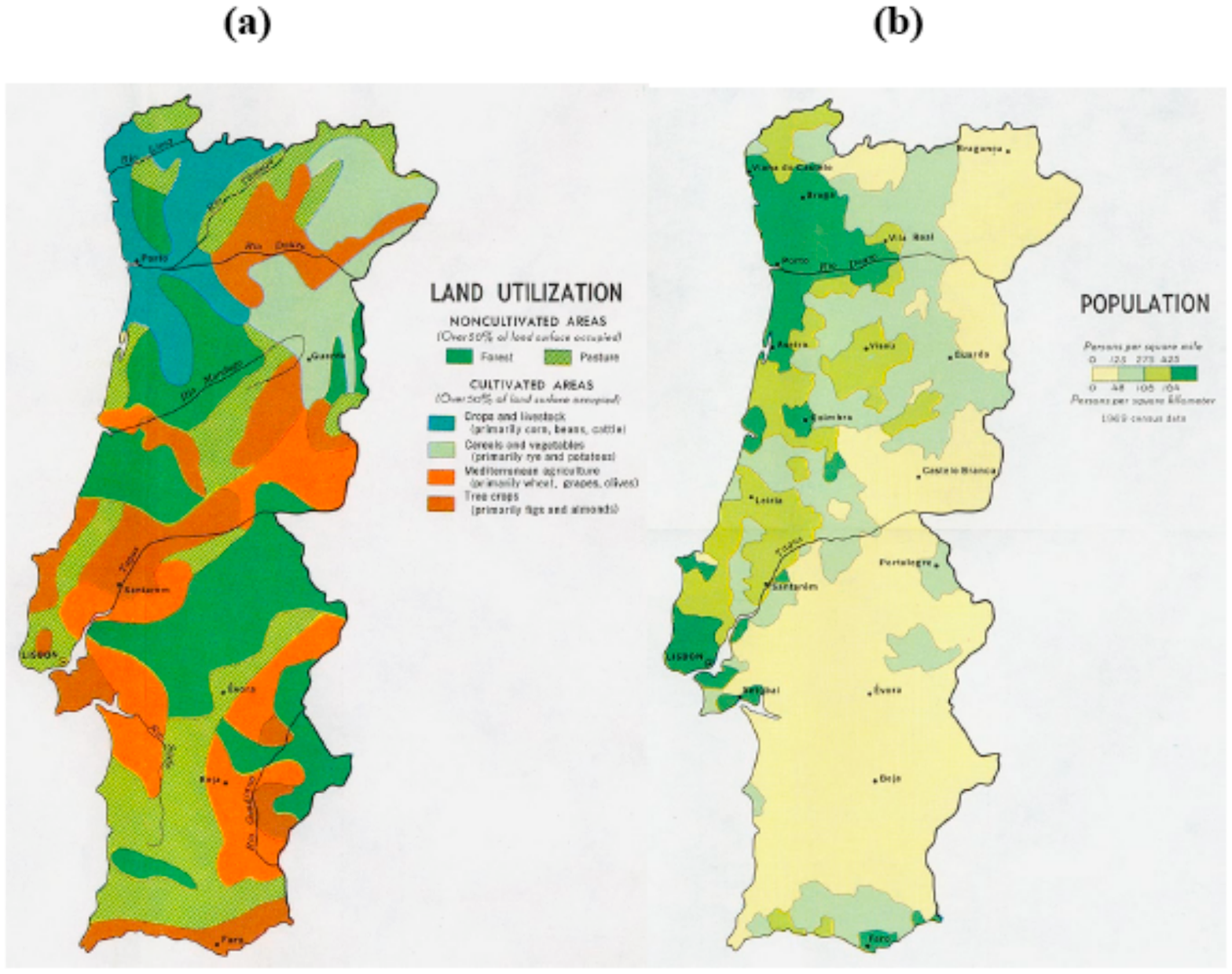}
\caption{Map of Portuguese land use panel (a) and population density panel (b). Produced by the Central Intelligence Agency and downloaded at www.lib.utexas.edu/maps/portugal.html .}
\label{port_map}
\end{center}
\end{figure}

Figure~ \ref{fig_birds} shows contour plots for the probability of a crested lark sighting for a single spline estimator and a mixture of splines estimators. This figure and the land use and population maps, reproduced in figure, \ref{port_map}  show why a mixture of at least 2 splines is necessary. Mainland Portugal is split by the river Tagus. The population and land use maps show that the population of Portugal is concentrated to the north of the river, in particular in the northwest where the land is given over to wine production. The interior of the north is dry and mountainous.  To the south of the river the predominant land form is rolling hills and the area is much less densely populated. The cultivated areas in the south are primarly for cork production but there are large tracts of forested areas.

Figure  \ref{fig_birds}(b) shows that in the southern part of Portugal the 
probability of sighting a crested lark  varies considerably and that these 
variations can occur abruptly.  These abrupt changes correspond to changes in 
the topography of southern Portugal.  The areas of high probability correspond 
to forested/tree crop areas or major rivers.  The areas of low probability 
correspond to pasturable lands.  
In contrast the probability of a crested lark sighting in the northern part of 
Portugal has little variation; the high population density together with the 
mountainous interior means that the probability is of  sighting is uniformly 
low. Thus the degree of smoothing required depends on the covariates $east$ 
and $north$.  Figure  \ref{fig_birds}  (a) shows that a single spline 
estimator cannot simultaneously capture the abrupt changes in southern 
Portugal and the smooth changes in the north.

\begin{comment}
\begin{figure}[bh]
\begin{center}
\includegraphics[scale=1]{roc&post_prob}
\end{center}
\caption{ROC for out of sample data (data) for different number of mixture components.}
\label{fig_roc}
\end{figure}

%Figure ??? shows the mixing function $\pi_{12}$ for a mixture with 2 components.
\begin{figure}[bh]
\begin{center}
\includegraphics[scale=.5]{1vsmix}
\end{center}
\caption{Contour plot for $\Pr (\mbox{crested lark sighting}=1|east, north)$ for a single spline estimator, panel (a) and a mixture of splines estimator panel (b).}
\label{fig_birds}
\end{figure}

\begin{figure}[htbp]
\begin{center}
\includegraphics[scale=.8]{portugal_maps}
\caption{Map of Portuguese land use panel (a) and population density panel (b). Produced by the Central Intelligence Agency and downloaded at www.lib.utexas.edu/maps/portugal.html .}
\label{port_map}
\end{center}
\end{figure}
%
\end{comment}

\begin{comment}
\begin{table}[ht!]
\begin{center}
\begin{tabular}{|l|l|}
\hline
Number of Components & Posterior Probability \\ \hline
1&0.00  \\
2&0.94 \\ 
3& 0.05\\
4&0.01\\ \hline
\end{tabular}
\end{center}
\caption{Posterior probability of the number of mixture components.}
\label{table_3}.
\end{table}
\end{comment}

\subsection{Probability of belonging to a union}\label{ss:union} 
This example shows how our methodology can be extended higher dimensions by  
modeling the probability of union membership as a function  of three 
continuous variables,  {\it years education}, {\it wage} and {\it age},  and 
three dummy variables, {\it south} (1=live in southern region of USA), {\it 
female} (1=female) and {\it married} (1=married).  
The data consists of 534 observations on US workers and can be found in Berndt 
(1991) and at http://lib.stat.cmu.edu/datasets/CPS\_85\_Wages. 
Ruppert, Wand and Carroll (2003) estimate the probability 
of union membership using a generalized additive model 
without interactions. We model the three dimensional 
surface of the continuous covariates and our results suggest that this more 
appropriate than an additive model. Our model for the probability of union 
membership is given by (\ref{e: mixture}) with  $x=$({\it years education, 
wage, age, south, female, married}). However, we modify the 
regression function  (\ref{eqn_2.2.1}) to
take into account that three of the covariates are dummy variables and should 
be excluded from the non-parametric component by writing 
\begin{align*} 
g_{jr} (x) & = \alpha_{jr}^\prime z + f_{jr}(x^*) 
\end{align*}
where $x^*=$ ({\it years education}, {\it wage},  {\it age}), {\it wage} is 
in US \$/hr and {\it age} is in years. %; $z = (south, female, married)$. 
The dependent variable is 1 if the worker belongs to a union and 0 otherwise. 
 
\begin{figure}[bh]
\begin{center}
\includegraphics{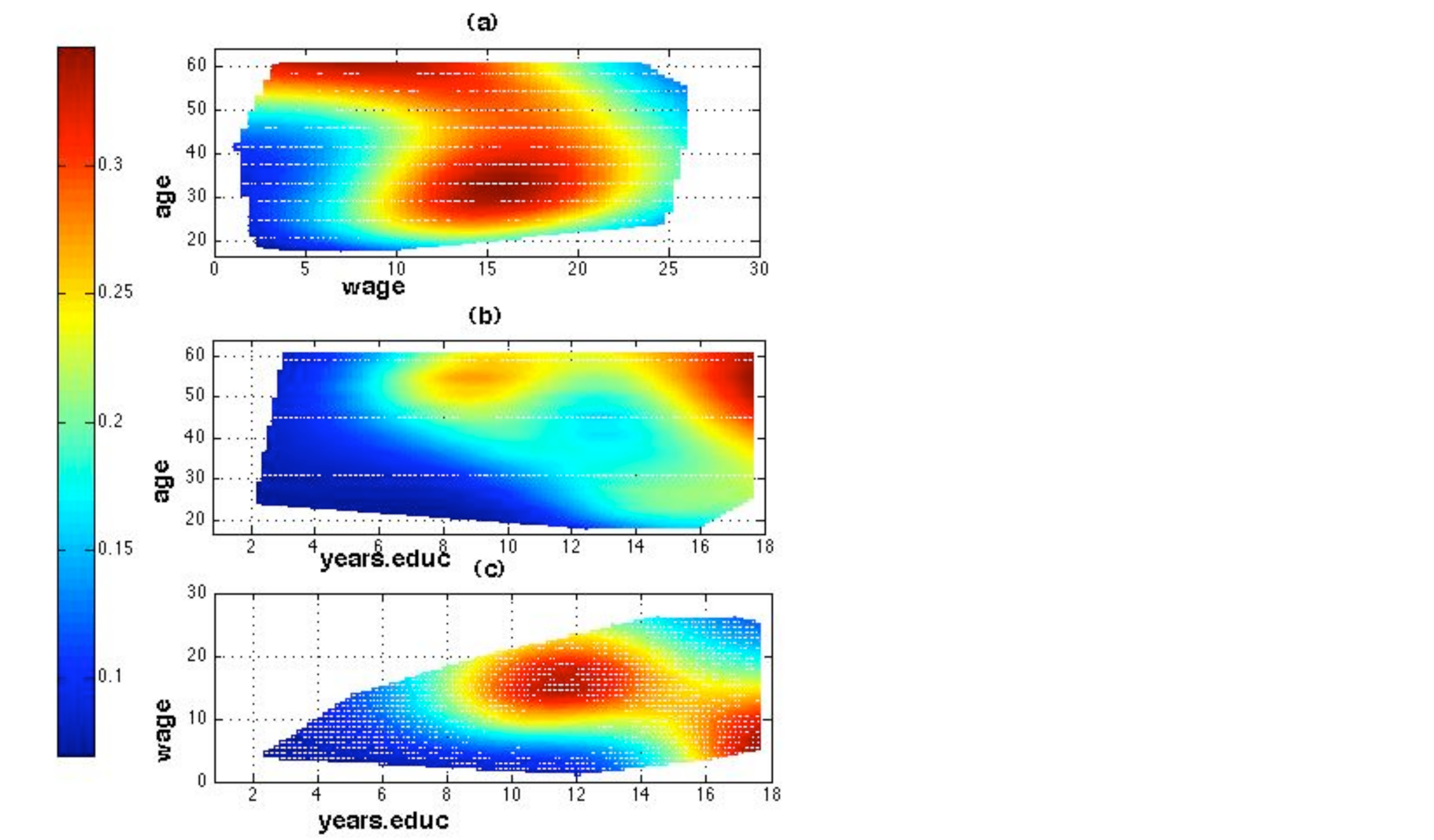}
\end{center}
\caption{Plot of $\Pr$(Union Member$|$ {\it wage, age}) at the mean {\it years education} panel~(a); $\Pr$(Union Member$|${\it age, years education}) at the mean {\it wage} panel~(b); $\Pr$(Union Member$|$ {\it wage, years education}) at the mean {\it age} panel~(c).  In all plots the dummy variables set at zero.}
\label{union}
\end{figure}

Our method chooses one component 100\% of the time. Figures ~\ref{union}~(a)~-~(c) show the joint marginal effect of two covariates at the mean of the third
one and setting the dummy variables to zero. 
These figures clearly show interactions among the continuous covariates.  For example figure ~\ref{union} (a)
shows that for workers whose {\it age} is less than 40, the probability of union membership initially increases with $wage$, before reaching a peak at a wage of about $\$15/hr$ and then declines.  For older workers this peak occurs at much lower wages, somewhere between $\$5$/hr and $\$10$/hr before declining sharply.  
Figure~\ref{union} (b) shows two modes.  For workers with an average {\it wage}, union membership peaks at 55 years and 8-10 {\it years education}. Interestingly union membership peaks again at 55 years and 18 {\it years education}, although this peak may be due to boundary effects. Figure ~\ref{union} (c) shows that for workers who did not finish high school ($< 12$ {\it years education}) the probability of belonging to union increases as {\it wage} increases. In contrast, for workers with some tertiary education($>14$ {\it years education}) the probability of belonging to a union is initially high and then decreases with increasing {\it wage}.
\begin{comment}
\begin{figure}[bh]
\begin{center}
\includegraphics{union}
\end{center}
\caption{Plot of $\Pr$(Union Member$|$ {\it wage, age}) at the mean {\it years education} panel~(a); $\Pr$(Union Member$|${\it age, years education}) at the mean {\it wage} panel~(b); $\Pr$(Union Member$|$ {\it wage, years education}) at the mean {\it age} panel~(c).  In all plots the dummy variables set at zero.}
\label{union}
\end{figure}
\end{comment}

\section*{Acknowledgment} We thank the referee, the associate editor and the editor for suggestions that improved the content and quality of the paper. We would especially like to thank Dr Tatyana Krivobokova for helping with the Adaptfit package and for carrying some of the computations for us. 
Sally Wood, Robert Kohn and Remy Cottet were supported by an ARC grant. 
%%%%%%%%%%%%%%%%%%%%% Appendix %%%%%%%%%%%%%%%%%%%%%%%%%%
\appendix
\myappendix{Sampling scheme} \label{appa}

%%%%%%%%%%%%%%%%%% section 3 % %%%%%%%%%%%%%%%%%%%%%%%%%
We estimate the binary regression probabilities
by their posterior means, with all unknown parameters and latent
variables integrated out. 
To make it easier to simulate from the posterior distribution
we introduce a number of latent variables
that are generated during the simulation and turn (\ref{e: mixture})
into a hierarchical model. The first is the number of
components $r$ at any point in the simulation.
Given $r$, define the vector of multinomial random variables
$\fgg _r  = \left\{ {\gamma _r(x_1), \ldots ,\gamma _r(x_n)}
\right\}$
 such that $\gamma _r (x_i)$ identifies the component in the mixture
that $w_i $ belongs to. We assume that 
\begin{align*} 
\Pr(\fgg _r |r,\fgd_r,x ) & = \prod_{i = 1}^n \Pr \{\gamma _r (x_i)|r,\fgd_r,x
\}\ , \quad \text{with} \quad 
\Pr\{ {\gamma _r (x_i) = j|\fgd _r }\} = \pi _{jr} (x_i )
\quad \text{and} \\
\Pr(\bfw|r,\fgg _r ,\bfg_r) & = \prod_{i = 1}^n \Pr\{w_i
|\gamma_r(x_i)=j,g_{jr}(x_i)\}
\end{align*}

To estimate the component splines, we follow 
Albert and Chib~(1993) and Wood and Kohn~(1998) and introduce a second level of
latent variables, $v_{ijr}$, also conditional on $r$, such that
\begin{align*}
v_{ijr}=z_i \alpha_{jr} + f_{jr}(x_i) + \epsilon_{ijr} \quad
\mbox{for}\quad j=1,\ldots, r \quad \mbox{and} \quad i=1, \ldots,
n,
\end{align*}
where $\epsilon_{ijr} \sim N(0,1)$. The latent variable $v_{jir}$
and the indicator variable $\gamma_{r}(x_i)$ are related to each
other and the observation $w_i$ by requiring that
\[
\begin{array}{lcrllcrllcl}
v_{ijr}  & >  & 0 & \mbox{if} & w_i & = & 1 & \mbox{and} &
\gamma_{r}(x_i) & = & j
\\
v_{ijr}  & <  & 0 & \mbox{if} & w_i & = & 0 & \mbox{and} &
\gamma_{r}(x_i) & = & j.
\end{array}
\]
If $\gamma_{r}(x_i) \ne j$, then $v_{ijr}$  is unconstrained.

The sampling scheme moves between models with differing numbers of components  
by using reversible jump \MCMC. 
To implement a reversible jump step to go from a model with $r$ components to 
a model with $r^\prime$ components it is necessary to have a proposal density 
in the $r^\prime$ component space. We form such proposals by first running 
separate \MCMC{} samplers for each $r = 1, \dots, R$ component models. This sampling scheme is described below under the heading of \lq Updating within a model.\rq

We now describe the complete sampling scheme. 
First, $r$ is initialized 
by drawing it from the prior $\Pr(r=j)=1/R, j=1, \dots, r$. 
Conditional on this value of $r$, we initialize
$\fgb_r$, $\fga_r$,$\fgt_r$, $\fgd_r$ by the posterior means of 
the iterates of the model with $r$ components. 
\begin{enumerate}
\item {\bf Moving Between Models} \newline \noindent
Let $X^c=(r^c,\Theta_r^c)$ be the current value of the parameters
in the chain, where
$\Theta_{r}^c=(\fga_{r}^c,\fgb_{r}^c,\fgd_{r}^c)$. 
We propose a new value of $X^p=(r^p,\Theta_r^p)$ and accept this
proposal using a Metropolis-Hastings (M-H) step. The M-H
probability of accepting such a proposal is
\begin{eqnarray}
\epsilon(X^c,X^p) & = & \min \left
\{1,\frac{p(X^p|\bfw)q(X^p,X^c)} {
p(X^c|\bfw)q(X^c,X^p)}\right \} \nonumber \\
 & = & \min \left
\{1,\frac{p(\bfw|X^p)p(X^p)q(X^p,X^c)} {
p(\bfw|X^c)p(X^c)q(X^c,X^p)}\right \} \label{eqn_a.1},
\end{eqnarray}
 where $q(X^c,X^p)$ is an arbitrary
transition probability function that moves the chain from $X^c$ to
$X^p$.

The proposal density $q(X^c,X^p)$ is given by $q(r^c \rightarrow
r^p)q(\Theta_{r^c}^c \rightarrow \Theta_{r^p}^p|r^p)$. That is, a
new value $r^p$ of $r$ is proposed and 
$\Theta_r^p$ is proposed conditional $r^p$. 
The value of $r^p$ is proposed as follows:
\begin{enumerate}
\item If $1<r^c<R$ then
$ q(r^c \rightarrow r^p=r^c \pm 1)  =  0.5 $
\item If $r^c=1$ then $q(r^c \rightarrow r^p=2) = 1$
\item If $r^c=R$ then $q(r^c \rightarrow r^p=R-1)=1$
\end{enumerate}
Then, conditional on $r^p$, we propose new values of the parameter
$\Theta_{r^p}^p=(\fgd^p_{r^p},\fgb^p_{r^p},\fga^p_{r^p})$ by doing the following.  To simplify notation, we write $r^p$ as $r$. 
\begin{enumerate}
 \item Draw $\fgd^p_{r}$ from
$MVT_5(\hat{\fgd},\hat{\Sigma}_{\fgd_r})$, where $\hat{\fgd}_r$ and 
$\hat{\Sigma}_{\fgd_r}$ are the sample mean and covariance of the iterates 
$\fgd_r^{[k]}$ from the individual MCMC scheme for a mixture of $r$ 
components. We use the notation $MVT_5(a,B)$ to denote a multivariate $t$ 
distribution with 5 degrees of freedom, location vector $a$ and scale matrix 
$B$. 

% At this stage the component dominance
%condition is checked. If this condition is not satisfied the
%proposed values are rejected. The new values of $r$,
%$\fgd_r$, $\fgb_r$, and $\fga_r$ are then
%set equal to their current values and we move to step~2. If these
%conditions are met then;
\item
Draw $\fgb^{*p}_{r}=(\fgb^p_r,\fga^p_r)$ from $MVT_5(\hat{\fgb}^*_r,\hat{\Sigma}_{\fgb^*_r})$.
 where $\hat{\fgb}^*_r $and $\hat{\Sigma}_{\fgb^*_r}$ are the sample mean and covariance of the iterates $\fgb_r^{*[k]}$ from the individual MCMC scheme for a mixture of $r$ components.  
 \end{enumerate}
The sampling scheme for a model of an $r$ component mixture is identical to the scheme for a within model move (step~2 below). 
\item {\bf Updating within Model}
 \newline \noindent
Given the new values of $r$, $\fgd_r$, and $\bfg_r$, the parameters specific 
to that model are updated as follows:
\begin{enumerate}
\item
Draw $\fgg_r, \bfV_r$ simultaneously from
$P(\fgg,\bfV_r|\bfw,\bfg_{r}, \fgd_r,\fgt_r)$ by first
drawing $\fgg_r$ and then drawing $\bfV_r$ conditional on the
value of $\fgg_r$.
\begin{enumerate}
\item
To draw $\fgg_r$ from $\Pr(\fgg_r| \bfw,\fgd_r,\bfg_r)$
note that
\[
\Pr(\fgg_r| \bfw,\fgd_r,\bfg_r)=\prod_{i=1}^n
\Pr \{\gamma_r(x_i)| w_i,\fgd_r,g_{jr}(x_i)\}.
\]
If $w_i=1$ then,
\begin{eqnarray*}
\Pr\{\gamma_r(x_i)=j|w_i=1,\fgd_r,\bfg_r\} & = &
\frac{\Pr\{w_i=1|\gamma_r(x_i)=j,\fgd_r,g_{jr}(x_i)\}
\Pr\{\gamma_r(x_i)=j|\fgd_r\}} {\sum_{k=1}^{r}
\Pr \{w_i=1|\gamma_r(x_i)=k,g_{kr}(x_i),\fgd_k\}
\Pr\{\gamma_r(x_i)=k|\fgd_r\}} \\
& = & \frac{\Phi\{g_{jr}(x_i)\}\Pr\{\gamma_r(x_i)=j|\fgd_r\}}{
\sum_{k=1}^r\Phi\{g_k(x_i)\}\Pr\{\gamma_r(x_i)=k|\fgd_r\}}.
\end{eqnarray*}

Similarly, if $w_i=0$ then,
\[
\Pr\{\gamma_r(x_i)=j|w_i=0,\fgd_r,\bfg_r\}=
\frac{[1-\Phi\{g_{jr}(x_i)\}]\Pr\{\gamma_r(x_i)=j|\fgd_r\}}{
\sum_{k=1}^r[1-\Phi\{g_{kr}(x_i)\}]\Pr\{\gamma_r(x_i)=k|\fgd_r\}}.
\]
\item
To draw $\bfV_r$ from $p(\bfV_r|\bfw,\fgg_r,\bfg_r )$ note
that,
\[
p(\bfV_r|,\bfw,\fgg_r,\bfg_r ) = \prod_{i=1}^n\prod_{j=1}^r
p(v_{ijr}|g_{jr}(x_i), \gamma_r(x_i),w_i).
\]
If $\gamma_r(x_i)=j$ and $w_i=1$, then $v_{ijr} \sim
N(g_{jr}(x_i),1)$, and is constrained to be positive.
\newline
\noindent If $\gamma_r(x_i) = j$ and $w_i=0$, then $v_{ijr} \sim
N(g_{jr}(x_i),1)$, and is constrained to be negative. \newline
\noindent If $\gamma_r(x_i)\ne j$ then draw $v_{ijr}$ from its
unconstrained distribution, which is $N(g_{jr}(x_i),1)$.
\end{enumerate}
\item
Draw $\fgd_r$ from $p(\fgd_r|\fgg_r)$ using a M-H
step. The conditional posterior distribution of $\fgd_r$ is

\begin{eqnarray}
p(\fgd_r|\bfw,\fgg_r) & = & p(\fgd_r|\fgg_r) \nonumber \\
& \propto & p(\fgg_r|\fgd_r)p(\fgd_r) \nonumber \\
& = & p(\fgd_r) \prod_{i=1}^n \frac{\exp\{\sum_{j=1}^r
\delta_{jr}z_i \gamma_r(x_i)\}}{\sum_{k=1}^r
\exp\{\delta_{kr}z_i\}} \label{eqn_a.2}
\end{eqnarray}
and $\fgd_r \sim N(0,10I)$. Our proposal density is
$MVT_5(\fgd_{max},\cal V_{\delta})$ where $\fgd_{max}$ is that
value of $\fgd_r$ that maximizes
$p(\fgd_r|\bfw,\fgg_r)$ in (\ref{eqn_a.2}), and
$\cal{V}_{\delta}$ is the negative of the inverse of the second
derivative of $\log\left[p(\fgd_r|\bfw,\fgg_r)\right]$. 
%The {\em
%component dominance} condition is then checked.  If this condition
%is not met then the newly drawn value of $\fgd_r$ is rejected.
\item
Draw $\fgb_r,\fga_r$ simultaneously from
       $p(\fgb_r,\fga_r|\bfV_r,\fgt_r)$ by first drawing $\fga_r$ and then
       conditional on this value of $\fga_r$ drawing
       $\fgb_r$:
       \begin{enumerate}
       \item
       To draw $\fga_r$ note that
       \[
       p(\fga_r|\bfV_r,\fgt_r) = \prod_{j=1}^r
       p(\alpha_{jr}|\bfv_j,\tau_{jr})
       \]
       and $p(\alpha_{jr}|\bfv_j,\tau_{jr})
        \sim N(\cal{M}_{\alpha},\cal {V}_{\alpha})$ where,
        \[
       \mbox{$\cal {V}$}_{\alpha}=[Z'Z-Z'X\tau_{jr}[\tau_{jr} X'X+I]^{-1}X'Z]^{-1}
       \]
       and
       \[
       \mbox{$\cal {M}$}_{\alpha}= \mbox{$\cal{V}$}_{\alpha}\left[Z'\bfv_{jr}-Z'X\tau_{jr}
       [\tau_{jr}X'X+I]^{-1}X'\bfv_{jr} \right].
       \]
       \item
       To draw $\fgb_r$ note that
       \[
       p(\fgb_r|\bfV_r, \fgt_r,\fga_r)
       =
     \prod_{j=1}^r p(\beta_{jr}|\alpha_{jr},\bfv_{jr},\tau_j)
      \]
      and $
      p(\beta_{jr}|\bfv_{jr},\tau_{jr}, \alpha_{jr}) \sim N(\cal {M}_{\beta}, \cal {V}_{\beta})$
      where
      \[
       \mbox{$\cal {V}$}_{\beta} =\tau_{jr}[\tau_{jr}X'X+I]^{-1}
       \] which is diagonal and
       \[
       \mbox{$\cal {M}$}_{\beta} =  \mbox{$\cal {V}$}_{\beta} X'\bfv_{jr}
       \]
      and $\bfv_{jr}=\bfv_{jr}-Z\alpha_{jr}$.
        \end{enumerate}
    \item
 Draw $\fgt_r$ simultaneously from
$p(\fgt_r|\fgb_r)$ by drawing from
         \[
        p(\fgt_r|\fgb_r)=\prod_{j=1}^r
        p(\tau_{jr}|\beta_{jr})
        \]
        and then re-label $\gamma_r(x_i)$ for $i=1,\ldots,n$ so that $\tau_{1r},> \ldots, >
        \tau_{rr}$ (see Stephens, 2000).
\end{enumerate}
\end{enumerate}

\myappendix{Constructing the design matrix} \label{appb}

This appendix outlines how we construct the design matrix $X$ to allow for a 
large number of observations and a moderate number of covariates. 

\begin{enumerate}
\item Normalize the values of all covariates to lie in the interval [0,1] 

\item Choose the number $m$ and  location of knots, so that in a given hypercube of width $\epsilon$, ($\epsilon$ is typically chosen to be 0.05)  and dimension $p$, a knot is placed at the centre of gravity of the hypercube.  If  there are no points in a hypercube then a knot is not chosen for that hypercube.  If the data are equally spaced across a grid where the grid length = $\epsilon$ then this results in $m=n$.  If the data are clustered, as is often the case in high dimensional data, then this technique results in $m <n$. 

\item Let $\tilde{x}_j$  be the position of the $j^{th}$ knot and let $x_i^*$ be the $i^{th}$ row of the normalized covariates. Radial basis functions, denoted by $\phi_{ij}$, are constructed such that $\phi_{ij}=||x_{i}^{\ast }-\tilde{x}_{j}||^{a}\ast \log (||x_{i}^{\ast
}-\tilde{x}_{j}||),$ $a=2\ast $ceil$(\frac{p}{2}+0.1)-p$,  where ceil$(x)$ means to round $x$ up to
the nearest integer value. This means that the $(i,j)^{th}$ element of the
 $n \times m$ design matrix $X$ is equal to $\phi_{ij}$ for $i=1,\ldots,n$ and $j=1,\ldots, m$. 
 \item
To limit  the dimension of the design matrix  we take a singular value decomposition of $X$, s.t. 
$X=U \Lambda V'$ where $U$ and $V'$ are square,
orthonormal matrices and $\Lambda$ is an $n\times m$ matrix, with nonnegative numbers on the diagonals, $\lambda_{ii}$  for $i=1,\ldots, m$ where $\lambda_{11} > \ldots  >\lambda_{mm}$, and zeros off the diagonal.  We then let $\lambda_{ii}=0$ for $i>l'$, where $l' =25$. We choose $l'$ in this way because typically  $1-\sum_{i=1}^{l'}{\lambda_{ii}^2}/\sum_{i=1}^m{\lambda_{ii}^2}< 1\times10^{-10}$. 
\item We re-form $X$ by letting $X=U\Lambda$. The design matrix $X$ is now a $n \times l'$ matrix. Note that $XX'=U\Lambda^2U'$ has the eigenvalue decomposition $QDQ'$, so that the resulting $n \times l'$ design matrix could have been formed by performing an eigenvalue decomposition on the $n\times n$ matrix  $XX'=QDQ'$ and setting $d_i=0$ for $i>l'$, however if $n$ is large performing an eigenvalue decomposition is computationally intractable.
\end{enumerate}

\noindent {\bf References}
\parindent=0pt

%\hangindent=0.2in \hangafter=1 Albert,A. \&  Anderson,
%J.A.~(1984). On the existence of maximum likelihood estimates in
%logistic regression models, {\em Biometrika}. {\bf 71},1--10.

 \hangindent=0.2in \hangafter=1
Albert, J. and Chib, S.~(1993). Bayesian analysis of binary and
polychotomous response data. {\em Journal of the American
Statistical Association}, {\bf 88}, 669--679.

\hangindent=0.2in \hangafter=1 Berndt E. R., The practice of Econometrics, New York: Addison-Wesley. (1991)

 \hangindent=0.2in \hangafter=1 Denison, D.G.T., Mallick,
B.K. \& Smith, A.F.M.~(1998). Automatic Bayesian curve fitting.
{\em Journal of the Royal Statistical Society B}, {\bf 60},
333--350.

\hangindent=0.2in \hangafter=1 Fan, J., Upadhye, S. and
Worster, A. (2006). Understanding receiver operating characteristic (ROC) 
curves, {\em Canadian Journal of Emergency Medicine}. 
{\bf 8}, 19-20. \\ 
http://www.caep.ca/template.asp?id=C4F7235436434ADAAB02D6B3E9C7A197

\hangindent=0.2in \hangafter=1 Friedman, J.H. and  Silverman, B.W.
(1989). Flexible parsimonious smoothing and additive modeling,
{\em Technometrics}. {\bf 31}, 3-39. 

\hangindent=0.2in \hangafter=1 Friedman, J.H.~(1991). Multivariate
adaptive regression splines (with discussion), {\em The Annals of
Statistics} {\bf 19}, 1-141.

\hangindent=0.2in \hangafter=1 Green, P.J.~(1995) Reversible jump
MCMC computation and Bayesian model determination. {\em
Biometrika}, {\bf 82}, 711-732.

\hangindent=0.2in \hangafter=1 Gu, C. (1992), 
 Cross-validating non-Gaussian data, {\em Journal of
Computational and Graphical Statistics}, {\bf 1}, 169-179.

\hangindent=0.2in \hangafter=1 Holmes, C.C. and Mallick, B.K. (2003)   Generalized Nonlinear Modeling With Multivariate Free-Knot Regression Splines,  {\em
Journal of the American Statistical Association,} {\bf 98}, 352-368.

\hangindent=0.2in \hangafter=1 Jacobs, R.A., Jordan, M.I., Nowlan,
S.J. and  Hinton, G.E.~(1991). Adaptive mixtures of local experts,
{\em Neural Computation} {\bf 3}, 79--87.

 \hangindent=0.2in
\hangafter=1 Jordan, M.I. and  Jacobs, R.A.~(1994). Hierarchical
mixtures-of-experts and the EM algorithm, {\em Neural Computation}
{\bf 6}, 181-214.

 \hangindent=0.2in
\hangafter=1 Krivobokova, T., Crainiceanu, C.M.  and Kauermann, C. (2006) 
Fast adaptive penalized splines. To appear in  {\em Journal of Computational and Graphical Statistics}.

\hangindent=0.2in \hangafter=1 Loader, C. (1999). {\em Local
regression and likelihood}, New York:~Springer

\hangindent=0.2in \hangafter=1 Luo, Z. and  Wahba, G.~(1997). Hybrid
adaptive splines, {\em Journal of the American Statistical
Association} {\bf 92}, 107-114.

\hangindent=0.2in \hangafter=1 McCullagh, P.  and   Nelder,
J.A.~(1989). {\em Generalized linear models (2nd edition)}, New
York:~Chapman Hall.

\hangindent=0.2in \hangafter=1
 Rao, C.R. (1973, {\em Linear statistical inference and it
 applications (2nd edition)}, New York: John Wiley.

%\hangindent=0.2in \hangafter=1
%Ripley  B.D.~(1996). {\em Pattern Recognition and Neural
%Networks}. Cambridge University Press.

\hangindent=0.2in \hangafter=1
 Ruppert D., Wand, M.P. and Carroll R.J. (2003). {\em Semiparametric Regression}.
 Cambridge University Press.
 
\hangafter=1 Smith, M. and  Kohn, R. (1996).
Nonparametric regression using Bayesian variable selection,  {\em
Journal of Econometrics} {\bf 75}, 317-344.

\hangindent=0.2in \hangafter=1 Stephens, M. (2000). Dealing with
label-switching in mixture models. {\em Journal of the Royal
Statistical Society B}, {\bf 62}, 795--809.

\hangindent=0.2in \hangafter=1 Tierney, L. (1994). Markov chains
for exploring posterior distributions (with discussion). {\it The
Annals of Statistics}, {\bf 22}, 1701-1762.

%\hangindent=0.2in \hangafter=1 Wahba, G.~(1990). {\em Spline
%Models for Observational Data}. SIAM, Philadelphia. CBMS-NSF
%Regional Conference Series in Applied Mathematics, Vol. 59.

%\hangindent=0.2in \hangafter=1 Wahba, G., Gu, C., Wang, Y. \&
%Chappell, R.~(1995). Soft classification, a. k. a. risk
%estimation, via penalized log likelihood and smoothing spline
%analysis of variance, in D. Wolpert, ed., {\em The Mathematics of
%Generalization, Santa Fe Institute Studies in the Science of
%Complexity, Proc. Vol XX,} Addison-Wesley, reading, MA, 329--360.

\hangindent=0.2in \hangafter=1 Wahba, G., Wang, Y., Gu, C., Klein,
R. and  Klein, B~(1997). ``Smoothing spline ANOVA for exponential
families, with application to the Wisconsin epidemiological study
of diabetic retinopathy'', {\it Annals of Statistics}, {\bf 23},
1865--1895.

\hangindent=0.2in \hangafter=1 Wang, Y.~(1994), Unpublished
doctoral thesis, University of Wisconsin-Madison.

\hangindent=0.2in \hangafter=1 Wang, Y.~(1997).  GRKPACK: Fitting
Smoothing Spline ANOVA Models for Exponential Families, {\em
Communications in Statistics: Simulation and Computation}, {\bf
26}, 765--782.

\hangindent=0.2in \hangafter=1 Wood, S.A. and  Kohn, R. ~(1998). A
Bayesian approach to robust binary non-parametric regression, {\em
Journal of the American Statistical Association,} {\bf 93},
203--213.

\hangindent=0.2in \hangafter=1 Wood, S.A., Jiang W. and  Tanner,
M.A.~(2002). Bayesian mixture of splines for spatially adaptive
nonparametric regression, {\em Biometrika}, {\bf 89}, 513-528.

\hangindent=0.2in \hangafter=1 Wood, S.A.,  Kohn, R.,  Jiang W. and 
Tanner, M.A.~(2005). Mixing on the outside, {\em Unpublished 
technical Report,  University of New South Wales}.

\hangindent=0.2in \hangafter=1 Wood, S.N. ~(2006). 
 {\em Generalized Additive Models: An Introduction with R}, Chapman Hall/CRC

%\hangindent=0.2in \hangafter=1 Wu, S.I., Chou, P. \& Tsai,
%S.T.~(2001). The impact of years since menopause on the
%development of impaired glucose tolerance, {\em Journal of
%Clinical Epidemiology}, {\bf 54}, 117-120.

%\hangindent=0.2in \hangafter=1 Yau,P., Kohn R. \& Wood,
%S.A.,~(2003). Bayesian variable selection and model averaging in
%high dimensional multinomial nonparametric regression, {\em
%Journal of Computational and Graphical Statistics}, {\bf 12},
%23-54.

%\bibliographystyle{asa}
%\bibliography{bin_me}

%%%%%%%%%%%%%%%%%%%%%%%%%%% FIGURES %%%%%%%%%%%%%%%%%%%%%%%%

\end{document}